\documentstyle[twoside,fleqn,espcrc2,epsfig]{article}

\newcommand{\AmS}{{\protect\the\textfont2
  A\kern-.1667em\lower.5ex\hbox{M}\kern-.125emS}}

\title{Dual vortices in Abelian projected SU(2) in the Polyakov gauge
\thanks{work partially supported by the Department of Energy under  
Grant No. DE-FG05-91ER40617}}

\author{K. Bernstein, G. Di Cecio and R.W. Haymaker
\address{Department of Physics and Astronomy, Louisiana State University,\\
Baton Rouge, Louisiana 70803 USA}}

\begin{document}

\begin{abstract}
We study dual Abrikosov vortices in Abelian projected SU(2) gauge theory 
in the Polyakov gauge. We show that vortices are present in this gauge but 
they are suppressed with respect to the maximal Abelian gauge. We interpret 
this difference in terms of the shielding of the electric charge by 
the charged coset fields.
\end{abstract}

\maketitle

\section{Introduction}
It is widely conjectured that confinement in $QCD$ can be understood in 
terms of dual superconductivity of the vacuum \cite{hay}. 
Inherent to this picture is 
the identification, by means of an Abelian projection \cite{hooft}, 
of a $U(1)$ subgroup 
defining the monopoles (the analogue of Cooper pairs in ordinary 
superconductivity) that would condense. In this sense an impressive number 
of results has been achieved in the maximal Abelian projection. 
In particular 
we want to mention that in this gauge the constraint effective potential as 
a function of the monopole field 
has been calculated and a symmetry breaking minimum has been found in the 
confining phase \cite{chern}. 
Moreover dual Abrikosov vortices have been explicitly  
seen and the parameters of dual superconductivity measured \cite{singh,cea}. 
One possible 
question is whether these results can be reproduced  with a different choice 
of the Abelian projection. In Ref. \cite{pisa}, a non-zero vacuum expectation 
value of a monopole field operator has been found in the Polyakov gauge in 
the confining phase, thereby signaling the spontaneous breaking of the $U(1)$ 
gauge symmetry. In the following we study dual Abrikosov vortices in the 
Polyakov gauge. Our aim is to establish 
a connection with the results of Ref. \cite{pisa} in this gauge  and to study the 
differences between the dual superconductivity parameters in the Polyakov  
gauge and in the maximal Abelian gauge. 

In Sec.~\ref{projection} we define the Abelian projection in the Polyakov 
gauge and discuss problems arising in the definition of static charges in 
the finite temperature case. In Sec.~\ref{measure} we present our results 
for the distributions of currents between static sources above and below 
the deconfining phase transition and make a comparison with similar results 
in the maximal Abelian gauge.

\section{Abelian projection in the Po\-lya\-kov gauge}\label{projection}
Let us write the Polyakov line at site $x$ as
\begin{eqnarray}
&&P(x) = \cos\theta_P(x)+i\phi(x)\sin\theta_P(x)\label{poly}\\
&&\phi=\hat{\phi}\cdot \vec{\sigma}\;\;\;\hat{\phi}\cdot\hat{\phi}=1
\;\;\;\theta_P\in[0,\pi]\nonumber
\end{eqnarray}
Diagonalization of $P(x)$ can be achieved by rotating  the vector $\hat{\phi}$ 
in the $\pm 3$ direction. One can then introduce in the standard way 
\cite{kron} Abelian links associated to a residual $U(1)$ invariance and 
doubly charged (with respect to this $U(1)$) coset fields. When fixing the 
gauge one has the freedom at every site to rotate $\hat{\phi}$ in the 
positive or negative $3-$direction. Translation invariance suggests we 
adopt the same rule at every site e.g. $\phi=+\sigma_3$ everywhere. 
With this prescription it is easy to show that the Abelian Polyakov line 
is given by $P_{ab}(x)=e^{i\theta_P(x)}$ and, since $\theta_P\in[0,\pi]$, 
$P_{ab}$ acquires a non-zero imaginary part due to this kinematical effect. 
This is different from the usual $U(1)$ case in which the angle associated 
to the Polyakov line is distributed symmetrically in the domain $(-\pi,\pi]$.
\begin{figure}[htp]
\epsfig{file=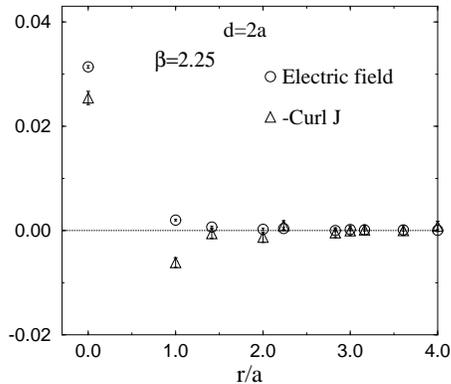,width=5.8truecm,angle=-90}
\label{fig1}
\caption{Electric field and curl of monopole currents in the confined phase 
for the Polyakov gauge.}
\end{figure}
However it can be shown \cite{preprint} that this is the only sensible 
prescription; if one tries to recover the usual $U(1)$ behavior of the 
Polyakov line by rotating $\hat{\phi}$ in the $+$ and $-$ direction with 
the same statistical weight then any correlation between Abelian Wilson loops 
vanishes. 
Since we want to represent a quark-antiquark source at finite 
temperature  with two Abelian 
Polyakov lines, $P_1P_2^\ast=e^{i(\theta_1-\theta_2)}$, we have to solve the 
problem that $P_1P_2^\ast$ 
 acquires a non-vanishing real part from the above described
 kinematical effect 
($\theta$ is distributed, according to the Haar measure, as $\sin^2\theta$).
Our solution is to define the quark-antiquark source as the connected part 
of the product of two Polyakov lines:
\begin{equation}
\langle P_1P_2\rangle -\langle P_1\rangle\langle P_2\rangle
\end{equation}
\begin{figure}[htp]
\epsfig{file=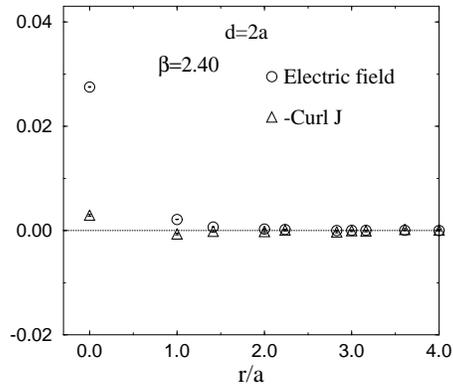,width=5.8truecm,angle=-90}
\label{fig2}
\caption{As in Fig. 1 but in the unconfined phase.}
\end{figure}

\section{Measurements and discussion}\label{measure}
We want to find direct evidence for the existence of dual Abrikosov 
vortices in the confined phase of the Abelian projected theory in the 
Polyakov gauge. Therefore we study the distribution of the electric field 
and of the monopole current around Abelian static sources, defined according
to the procedure described in the previous section. Our aim is to interpret 
the results in terms of the dual superconductivity parameters i.e. the dual 
coherence length $\xi_d$ and the dual London penetration depth $\lambda_d$.
We take our measurements at finite temperature on a $12^3\times 4$ lattice.
The longitudinal electric field and the curl of the magnetic monopole 
current between static charges  are defined as in \cite{singh}. 
In Fig.~1 we show the results in the confined phase ($\beta 
=2.25$) in the case of two sources separated by a distance $d=2$ in lattice 
spacings. We see a clear signal for both the electric field and the curl 
of the monopole currents on the axis of the two charges. As we move in the 
transverse direction $r$, we observe a rapid fall-off of the electric field 
while the monopole current shows a behavior consistent with the prediction 
of the dual Ginzburg-Landau theory. In Fig.~2 we show the results 
in the unconfined phase ($\beta=2.40$). In this case, the magnetic currents 
are much smaller and are consistent with zero in the region away from the 
axis of the two charges; moreover the electric field approaches zero less 
rapidly altough not clearly evident from the figure.
Our conclusion is that dual vortices are present in the confined phase 
and disappear in the unconfined phase; hence we bring support to the 
dual superconductivity scenario in the Polyakov gauge \cite{pisa}.

\begin{figure}[htp]
\epsfig{file=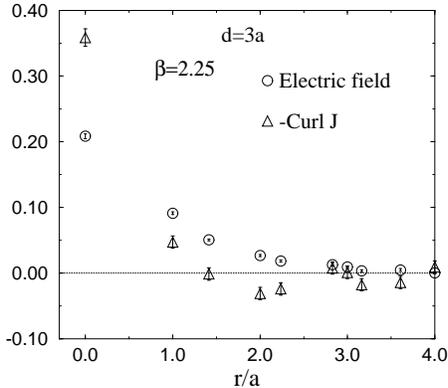,width=5.8truecm,angle=-90}
\label{fig3}
\caption{Electric field and monopole currents in the maximal Abelian gauge.}
\end{figure}

As already mentioned, similar results have been reported for the maximal 
Abelian gauge \cite{singh,cea}. 
In Fig.~3 we show the results in this gauge for 
the confined phase and $d=3a$. There are two main differences between 
these results and the correspondent results in the Polyakov gauge. First, 
the data in the maximal Abelian gauge support a non vanishing value for 
the coherence length $\xi_d$, while there is no evidence for it being 
different from zero in the Polyakov gauge. Then, the peak values of the 
electric field and the magnetic currents are more than an order of 
magnitude smaller in the Polyakov gauge than the correspondent values in 
the maximal Abelian gauge. We interpret these differences in terms of 
a different role played by the charged coset fields in the two projections. 
Specifically, we measure the spatial  distribution of the electric 
charge in the region between two static sources. This can be done 
measuring the divergence of the electric field defined, at site $x$, as
\begin{equation}
\sum_{i=1}^3
\frac{\langle\sin\theta_{PP^\dagger}(\sin\theta_{i4}(x)-
\sin\theta_{i4}(x-i))\rangle}
{a^3 e\langle\cos\theta_{PP^\dagger}\rangle}
\end{equation}
\begin{figure}[htp]
\epsfig{file=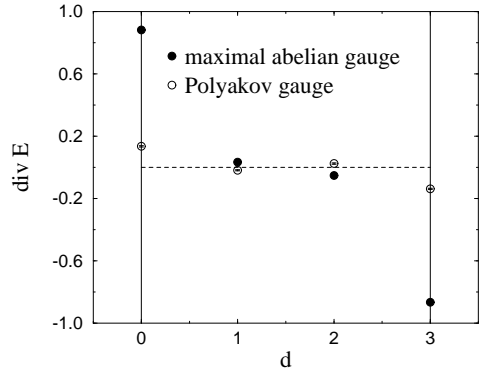,width=5.5truecm,angle=-90}
\label{fig4}
\caption{Divergence of the electric field along the axis connecting two 
static charges.}
\end{figure}
In Fig.~4 we show the results for two Polyakov lines separated 
by three lattice spacings. The figure clearly shows that the effective 
static charge is much smaller in the Polyakov gauge and that for 
$d\neq 0$ the coset fields in 
the two prescriptions respond in opposite ways to the presence of an 
electric charge.

Our conclusion is that dual vortices are present in the Polyakov gauge 
but are suppressed with respect to the maximal Abelian gauge due 
to the charged coset fields that are far more effective  in shielding 
the electric charge in the case of the Polyakov gauge.


\begin{thebibliography}{99}
%
\bibitem{hay} For a review  and further references see e.g. R.W. Haymaker, 
lectures given at the International School of Physics ``E. Fermi'', 
Varenna, 1995, hep-lat/9510035.
%
\bibitem{hooft} G. 't Hooft, Nucl. Phys. B 190 (1981) 455.
%
\bibitem{chern} M.N. Chernodub, M.I. Polikarpov and A.I. Ve\-se\-lov, 
preprint ITEP-TH-14-95.
%
\bibitem{singh} V. Singh, D.A. Browne and R.W. Haymaker, Phys. Lett. 
B 306 (1993) 115.
%
\bibitem{cea} P. Cea and L. Cosmai, Phys. Rev. D 52 (1995) 5152.
%
\bibitem{pisa} L. Del Debbio, A. Di Giacomo, G. Paffuti and P. Pieri, 
Phys. Lett. B 355 (1995) 255.
%
\bibitem{kron} A.S. Kronfeld, G. Schierholz and U.J. Wiese, Nucl.Phys. 
B 293 (1987) 461.
%
\bibitem{preprint} K. Bernstein, G. Di Cecio and R.W. Haymaker, 
preprint LSUHE No. 213-1996. 

\end{thebibliography}
\end{document}